# What Modern Vision Science Reveals About the Awareness Puzzle:

Summary-statistic encoding plus decision limits underlie the richness of visual perception and its quirky failures

*VSS Consciousness Symposium 2017*
*Ruth Rosenholtz, MIT Dept. of Brain & Cognitive Sciences, CSAIL*

## Introduction

There is a fundamental puzzle in understanding our awareness of the visual world. On one hand, our subjective experience is one of a rich visual world, which we perceive effortlessly. However, when we actually test perception, observers know surprisingly little (Figure 1). A number of tasks, from search, through inattentional blindness, to change blindness, suggest that there is surprisingly little awareness or perception without attention (e.g. Treisman and Gelade, 1980; Mack and Rock, 1998; Rensink, O'Regan, and Clark, 1997). Meanwhile, another set of tasks, such as multiple object tracking (MOT), dual-task performance, and visual working memory (VWM) tasks suggest that both attention and working memory have low capacity (e.g. Scimeca and Franconeri, 2015; VanRullen, Reddy, and Koch, 2004; Luck and Vogel, 2013). These two components together – poor perception without attention, and greatly limited capacity for attention and memory – imply that perception is impoverished. Throughout, I refer to this combination of rich subjective experience and poor task performance as "the awareness puzzle," though of course it is far from the only puzzle when it comes to understanding awareness.

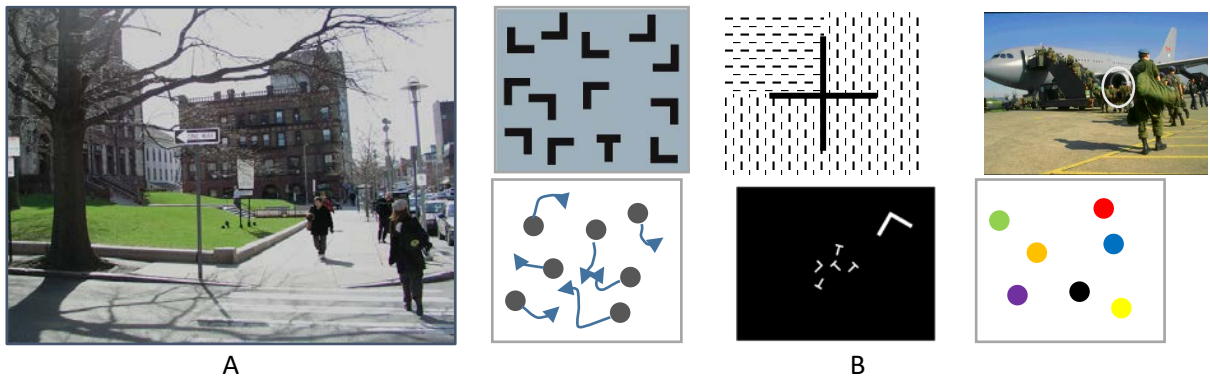

**Figure 1. The awareness puzzle. A.** Our subjective impressions of a real scene appear rich and full of detail. **B.** However, when tested in the lab, observers know surprisingly little. Visual search (top left), inattentional blindness (top middle), and change blindness (top right) all imply poor perception without attention. A number of other tasks (bottom: multiple object tracking, dual-task experiments, and visual working memory tasks) suggest that both attention and memory have low capacity.

How can we make sense of this awareness puzzle, of the riddle of our rich subjective experience coupled with poor performance on experimental tasks? Philosophers and researchers have offered two main theories. The first, here referred to as the "illusion theory", suggests that the rich subjective impression is merely an illusion, and therefore not incompatible with the impoverished perception observed in behavioral experiments. (Blackmore et al., 1995; Dennett, 1991, 1998; O'Regan, 1992; Rensink et al., 1997).

The second theory, here referred to as the "inaccessible theory" suggests that we are perversely aware of more than we can act upon (Block, 2011; Lamme, 2010). The rich percept is real, but the information is inaccessible when it comes to making decisions or otherwise taking action.

I will suggest that both of these theories have pinpointed interesting parts of the problem, but have not provided the solution to the awareness puzzle. My main claim is that, looked at in the right way, there is in fact no awareness puzzle. In particular, I will argue that the tasks that show limits are inherently difficult tasks, and that there exists a unified explanation for both the rich subjective experience and the apparent limits. Everyone in this symposium (Vision Sciences Society, St. Petersburg, FL, May 2017) has suggested some aspects of this story (Balas, Nakano, and Rosenholtz, 2009; Haberman and Whitney, 2009, 2011; Rosenholtz, 2011; Rosenholtz, Huang, and Ehinger, 2012; Cohen, Dennett, and Kanwisher, 2016; Leib, Kosovicheva, and Whitney, 2016; Rosenholtz, 2017), as have others in vision science (Ariely, 2001; Oliva & Torralba, 2006; Alvarez, 2011). Here I clarify previously suggested aspects of the solution, and suggest other key components for understanding the awareness puzzle.

## Revisiting the evidence for impoverished vision, part I

So, let us begin by discussing the evidence for impoverished vision. In particular, I begin by discussing visual search and change blindness. I leave discussion of MOT, VWM, and related phenomena to a later section, as I will argue that those phenomena predominantly expose different factors than search and change blindness. I have recently discussed inattentional blindness and dual-task performance in more detail (Rosenholtz, 2017), and therefore while I describe their place in the story I largely leave those details to that paper.

### Visual search may not provide evidence for impoverished vision

In the traditional view of visual search, search experiments probe limits of attention. Some amount of processing can occur "preattentively" in parallel across the visual field. Then, at some stage of processing, attention serially selects some portion of the input for higher-level processing. By comparing conditions that lead to difficult vs. easy visual search, we can supposedly determine at what stage selection occurs, and what processing is preattentive.

Experiments have generally shown that search is difficult whenever distinguishing the search target from other distractor items requires more than a simple basic feature like color or motion. This implies that only basic features – often referred to as "feature maps" – can be computed preattentively, and selection occurs early in visual processing. Because attention is a limited resource, this implies that vision is highly impoverished.

However, my lab has argued that we need to rethink the logic of search experiments and their implications for attention and the richness of perception (Rosenholtz et al., 2012ab; Zhang et al., 2015). We have shown that peripheral discriminability of a target-present from a target-absent patch predicts search performance (Figure 2). This suggests that search primarily pinpoints loss of information in peripheral vision, not attentional limits nor limits of preattentive processing.

One might ask why this distinction matters, since in both cases there would seem to be a loss of information, whether from the regions not fixated or from unattended regions, and therefore at first glance either system would appear to suggest impoverished vision. However, a peripheral vision explanation implies much less impoverished vision than the attentional explanation. In the attention explanation, unselected stimuli receive no processing after the bottleneck of attention. This means that many, if not most, tasks are impossible without attention. Peripheral vision, on the other hand, has simply lost information that happens to make difficult search hard. However, it also preserves a great deal of information (as we have proven, and will discuss in a later subsection). Most crucially, processing continues! One cannot recover the lost information without an eye movement, but the information that remains supports performance of many tasks, from guiding eye movements, through object recognition, to getting the gist of a scene and navigating the world.

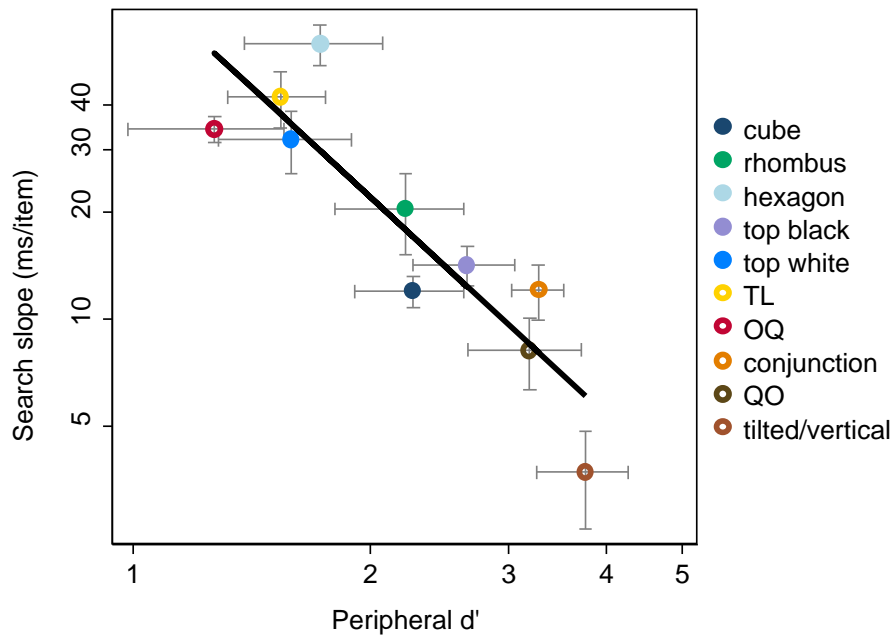

**Figure 2. Search performance vs. peripheral discriminability.** For a range of classic search conditions, greater peripheral discriminability d' leads to more efficient search, i.e. lower search slopes.

### Change blindness may not provide evidence for impoverished vision

Change blindness refers to the difficulty detecting a change to an image. In the lab, the experimental paradigm often involves flickering between two versions of an image, while introducing a brief blank frame between the pair in order to disrupt motion cues. The phenomenon itself is essentially the same as the childhood puzzle in which one must find the differences between two side-by-side images.

Many researchers have viewed change blindness as probing the limits of perception without attention. (Researchers have also offered a memory explanation for change blindness, of less relevance for the present discussion.) Supposedly, the observer manipulates a spotlight of attention, and perception is richer within that spotlight than outside of it. The difficulty of detecting a change implies that little perception occurs without attention.

However, peripheral vision is a factor in change blindness. We first categorized standard change blindness stimuli as easy, medium and hard, based on time to detect the changes (Figure 3). We then measured difficulty detecting a known and presumably attended change using peripheral vision. We found that for the hard changes, observers needed to fixate significantly closer to the change in order to perceive it, even though they knew in advance the identity of the change and its location (Sharan et al., under review).

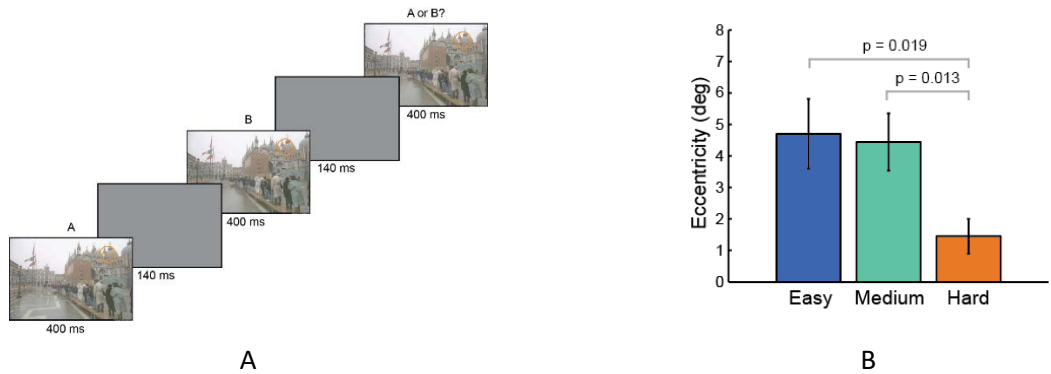

**Figure 3. Peripheral vision is a factor in change blindness. A.** Observers discriminated known changes in an A-B-X paradigm that requires them to identify whether the final image matches the first or the second one in the sequence. Fixation was enforced at various distances to the change. Orange circle shows one such fixation. The difference for this sequence was in the pattern on the ground. **B.** The threshold eccentricity (distance to the change) for easy-, medium-, and hard-to-detect changes. Harder changes require closer fixation to be discriminated.

Once again, a peripheral vision explanation implies that perception is richer than previously thought. The information lost in peripheral vision just happens to make it hard to detect changes. (This may not be due to chance. Researchers tend to design experiments with "interesting" changes, and the changes dubbed "uninteresting" may be easier to detect in peripheral vision.) However, peripheral vision preserves a great deal of information, which could support a wide range of tasks. Again, I discuss this in more detail in the following subsection.

### Peripheral vision encoding preserves a great deal of information

We have argued elsewhere, for roughly the last decade, that peripheral vision encodes its inputs in terms of a rich set of image statistics. These statistics are "summary statistics", meaning they pool information over sizeable local regions that grow with the distance to the point of fixation, i.e. the eccentricity. We have discussed our candidate image statistics elsewhere, and so do not go into detail here. This encoding leads to significant loss of information, and we have accumulated extensive evidence that this loss of information can predict difficulty recognizing peripheral objects, particularly in cluttered displays or scenes (Balas et al., 2009; Rosenholtz et al., 2012b; Zhang et al., 2015; Chang & Rosenholtz, 2016; Keshvari & Rosenholtz, 2016). The loss of information predicts difficult search conditions, while preserving the information necessary to predict easy "popout" search (Rosenholtz et al., 2012b; Zhang et al., 2015; Chang & Rosenholtz, 2016).

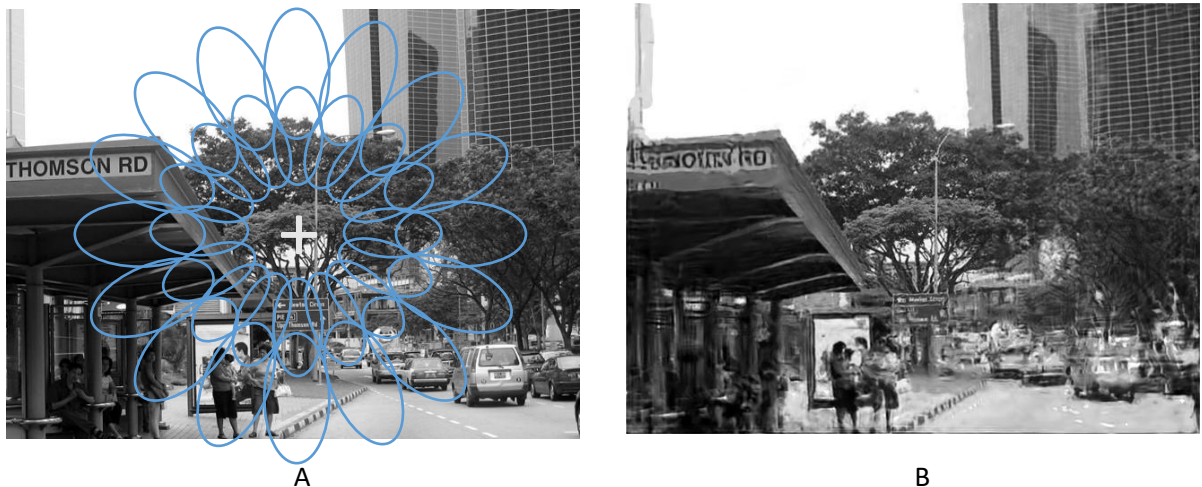

**Figure 4. Information encoded in a rich set of image statistics. A.** Original image, candidate pooling regions superimposed. They grow linearly with eccentricity. **B.** Image synthesized to have the same local image statistics as the original. This encoding captures a great deal of information, though some of the details of the image are unclear.

Nonetheless, this encoding preserves a great deal of information. It preserves sufficient information to predict human performance getting the gist of the scene, including scene category, upcoming turns, presence of a particular object like an animal or a stop sign, and what city appears in the photograph (Rosenholtz et al., 2012a; Ehinger & Rosenholtz, 2016). To get a sense of what information is encoded by a rich set of image statistics such as those proposed, one can synthesize images that contain the same statistics but are otherwise random (Rosenholtz, 2011; Freeman & Simoncelli, 2011; Rosenholtz et al., 2012a; Ehinger & Rosenholtz, 2016). As shown in Figure 4, this encoding preserves a great deal of information. That it does so is not surprising, as this scheme involves a large number of image statistics; as many as 1000 per pooling region. While we know surprisingly little about the information available in our rich subjective impression of the world, it seems plausible that this encoding scheme has sufficient information to support that subjective impression.

Examining Figure 4, however, it is clear that the encoding does not preserve certain details. One cannot read the Thomson Rd. sign, nor easily discriminate the number and types of vehicles. This ambiguity of the details could underlie poor performance in change detection experiments. Figure 5 shows a demo of this same synthesis technique applied to a pair of change detection images. When fixating 5 degrees away from the change, the model predicts difficulty detecting the change. However when fixating 1 degree away, the change becomes clear, in agreement with our data on change discrimination in the periphery (Sharan et al., under review).

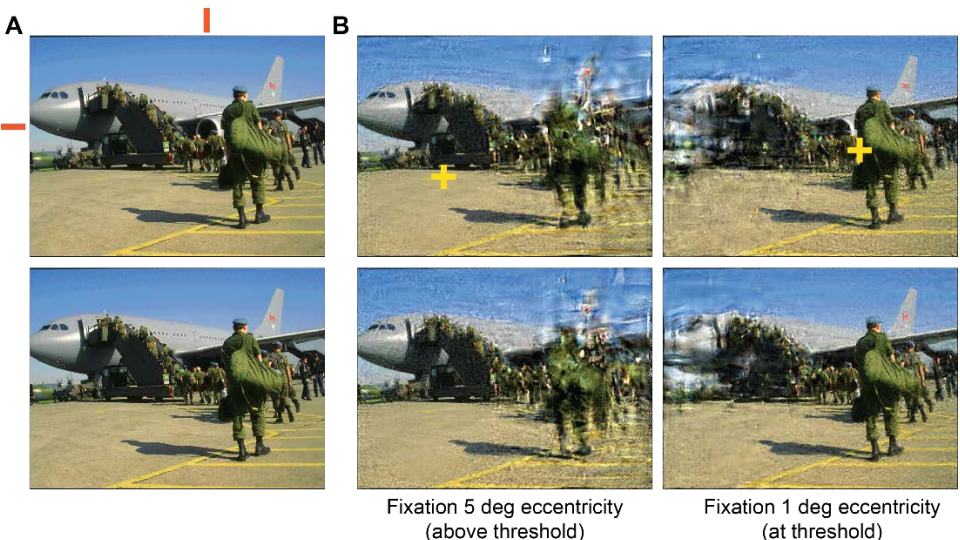

**Figure 5. Summary image statistics lose information about the details, which can lead to difficult change detection. A.** Image pair. Red bars indicate changed region. **B.** Synthesis visualizes the information available in a summary statistic encoding for a fixation 5 degrees (left) and 1 degree (right) from the change. Note that the change is clear in the latter pair, but not the former.

Humans are also quite good at reporting the ensemble properties of a set of objects. This includes low-level properties like the mean and variance of orientation or size, as well as higher-level properties such as mean facial expression. According to traditional theories of vision, such performance is surprising, as how could observers report the mean when extracting features of individual items requires attention. In fact, observers are poor at reporting the features of a particular item. However, in opposition to such theories, it is worth noting that no studies, to my knowledge, have found that observers are better at reporting the mean than one would expect from their ability to report whether a given feature was present anywhere in the display. In other words, observers have difficulty reporting the identity of a particular post-cued item, but seem to have access to something like a histogram of the features in the display.

As shown in Figure 6, a rich set of image statistics preserves sufficient information to support a wide variety of ensemble perception tasks. Sizes and orientations of items are largely preserved, but location information is lost; the lost information perhaps partially explains the difficulty reporting a particular item's identity. In addition, one can tell that the items on the right are oriented dark bars on a light background, and those on the left are black and white quadrisected disks against a gray background.

Authors often confuse image statistics with ensemble statistics of a set of items, so it is worth noting that the large set of image statistics can support perception of scenes and sets, but research has not demonstrated that a handful of ensemble statistics can support rich scene perception.

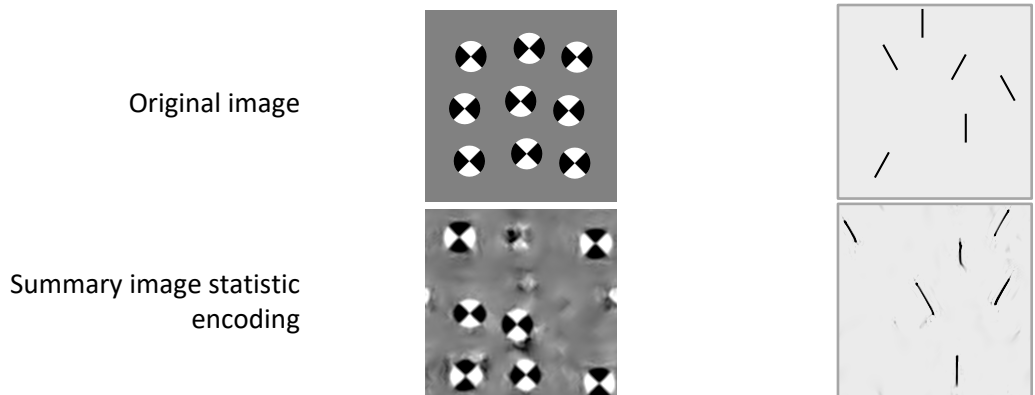

Original image

Summary image statistic encoding

**Figure 6. Summary statistic encoding can underlie the richness of ensemble perception.** These syntheses visualizing the encoded information rely on information from only a single pooling region, leading to greater loss of location information than with the full version of the model.

## Revisiting the evidence for impoverished vision, part II

It should already seem plausible from this discussion that there is no awareness puzzle. The same encoding of the visual input seems likely to predict the difference between easy and difficult visual search, the difference between easy and difficult change detection, to predict perception of the gist of scenes and sets of items, and to preserve enough information to support a rich subjective experience of the visual world.

However, Figure 1 referred to other perceptual phenomena not readily explained by a summary statistic encoding. These include inattentional blindness, the difficulty in performing two tasks at once, MOT, and the limited capacity of visual working memory. Though many of these tasks may, in part, be explained by a summary statistic encoding in peripheral vision, it is clear for most of these tasks that that cannot be the entire explanation. Encoding losses in peripheral vision particularly affect vision in cluttered or complex displays, which certainly applies to typical displays for these tasks. However, researchers have in some cases controlled for fixation and clutter and still found poor performance at these tasks.

I argue that this second group of tasks is inherently difficult. Consider, as a canonical example, a typical VWM task. An observer is shown an array of $k$ items, such as colored disks. After viewing this array for some time, the experimenter then shows another array. This array either duplicates the original, or differs, often in the color of one of the $k$ disks. In the traditional way of thinking of this task, the observer has $n$ memory slots to fill with features from each of the $k$ disks. Based on how performance varies as a function of the number of items in the display, one can supposedly infer the number of slots. Based on this logic, researchers have concluded that a typical observer has only around 4 slots; this suggests a very limited capacity for VWM.

This slots framework, however, makes many assumptions about how VWM works. If we think of the task at a more basic level, the observer must discriminate between the array to be remembered and all other arrays with at least one change. One could imagine that this discrimination would require a fairly complicated classifier. Just how complicated would depend upon the feature space on which the classifier operates. The feature space seems unlikely to be a vector of $k$ colors. Put another way, the arrays of colored disks likely occupy a very small subset of "perceptual space" – the images one is likely to see, represented in whatever high-dimensional encoding used by the visual system. Discriminating between such similar images might be quite difficult.

A very similar story applies to tasks such as reporting a post-cued member of an ensemble. This task is essentially a version of the VWM task, and is likely hard for the same reason. MOT tasks are inherently difficult for a somewhat different reason. Again, in the traditional interpretation, the visual system has $m$ attentional spotlights to deploy. Based on performance one can infer $m$, and as $m$ is low, one concludes that attention has limited capacity. However, as with VWM, this account makes strong assumptions about the mechanisms involved. At a more basic level, if the observer must track $k$ of $n$ items, then on each frame they must distinguish the actual $k$ targets from $n$-choose-$k$ other possible combinations of targets. In the case of tracking 4 of 9 items, for instance, the observer must distinguish the actual 4 targets from 125 other possibilities! One might again imagine that this task is inherently difficult in the abstract, though of course motion cues can make the task more tractable.

The inherent difficulty of these tasks, almost independent of the underlying representation, suggests that they target decision-level mechanisms. I have recently made a similar argument that inattentional blindness, dual-task performance, and likely search and change blindness may all encounter decision-level limits as well as encoding limits, and point the reader to that recent paper rather than reiterating the arguments here (Rosenholtz, 2017).

## The promise of a unifying explanation, and thoughts on a remaining puzzle

I have suggested, then, the following general mechanism underlying perception and awareness (Figure 7). First, a number of stages of processing compute a perceptual encoding of the visual input. Losses in this encoding lead to poor performance on a number of visual tasks (difficult search, change blindness), while preserving sufficient information to make other tasks relatively easy (easy search, and getting the gist of a scene or set), and perhaps to support our rich subjective percept of the world. Information may be lost at a number of the stages of perceptual processing. A particularly interesting loss appears to be an encoding in terms of a rich set of summary image statistics, derived from pooling image measurements across sizeable regions of the visual field, which grow with distance from the point of gaze. My lab has demonstrated that this loss alone explains a surprising number of visual phenomena. Nonetheless, processing continues beyond that stage of processing, and to the extent that necessary information survives, the observer can perceive groups, reason about materials and shape, recognize objects, and get the gist of a scene. However, ultimately performance on a task depends upon decision (inference) processes operating on the perceptual encoding. Some tasks (MOT, VWM, dual-task, etc.) may be difficult because they are inherently difficult given the encoding, and/or the decision processes may themselves have limits, e.g. on the allowable complexity of the inference (Rosenholtz, 2017).

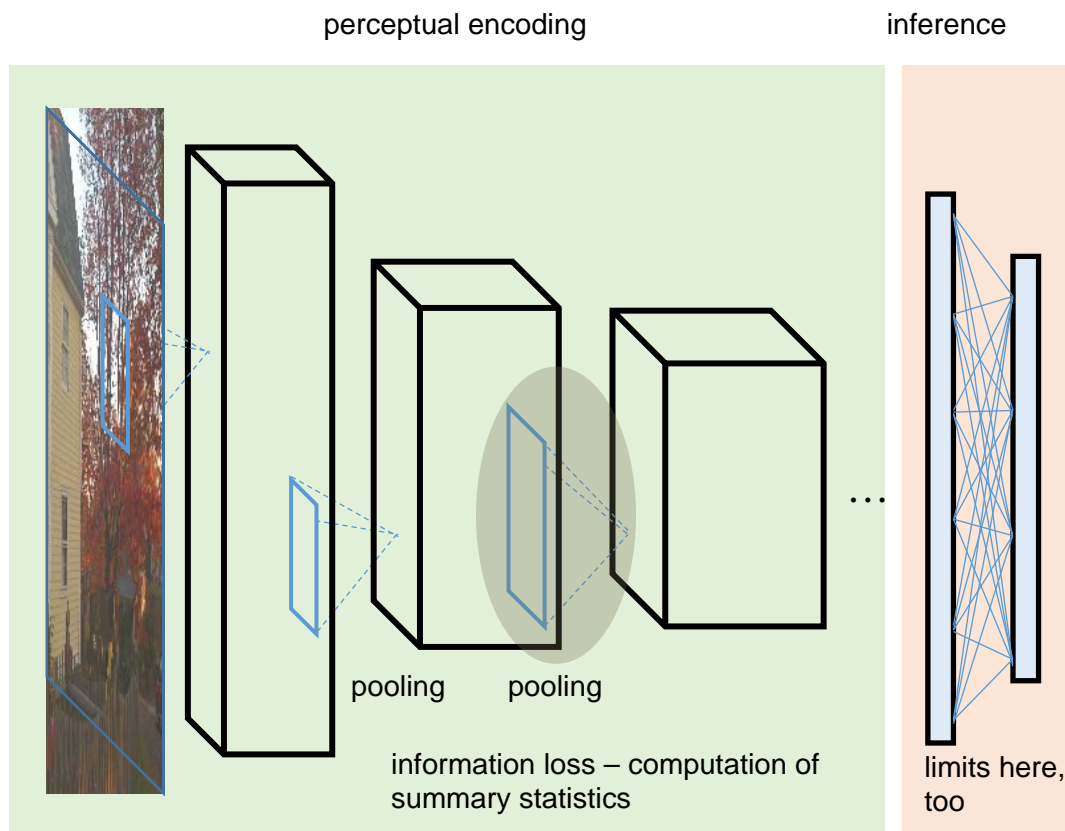

**Figure 7. Hypothesized architecture underlying "awareness puzzle" phenomena.** See text.

If there exists a single model that predicts the phenomena underlying the puzzle of awareness then there is no puzzle. The tasks that appear to show limits are simply hard tasks, either due to the encoding or due to limits on inference processes. The same system may predict that a great deal of information is available to support a rich subjective experience.

While the discussion so far seems promising, I have tricked the reader a bit; a puzzle remains. I have said that the summary statistic encoding preserves sufficient information to provide a rich subjective impression. However, presumably this subjective impression, too, derives from the visual system performing some task, which in turn is subject to limits on the inference processes. Does that inference process, too, support a rich subjective impression? Put another way, why is the "task" that leads to a rich subjective impression somehow easier than the task of tracking more than 4 objects?

This remaining puzzle sounds disturbingly reminiscent of the "inaccessible theory" of visual awareness: why is it that when the experimenter does not probe the observer with a task, the "inference" of a rich percept is easy to acquire, and yet when there is an experimenter-defined task (MOT, VWM, etc.), the task encounters limits? It could be that there is a distinction between having a concrete task and simply getting an impression of the scene – declarative vs. episodic perception, to parallel a similar distinction between different types of memory. However, having come this far, I think we do not have to give up quite yet.

Suppose we hypothesize that one can operationalize the task of getting a subjective impression in the following way: The visual system attempts to answer the question, "where does this stimulus lie in perceptual encoding space?" If the visual system lacks precision in answering this question, either due to encoding or inference limits, then this will lead to uncertainty about the stimulus. To get a sense of what this uncertainty might mean, we performed a mini-experiment as follows:

We took as our candidate perceptual encoding the last representational layer (the "last fully connected layer") of a convolutional neural network (CNN). CNNs have recently become very popular, as for the first time they allow computer vision to approach human performance on certain proscribed visual tasks. Researchers have also shown certain similarities between the representations learned by CNNs and those found in monkey physiology (Yamins et al., 2014). We used a CNN known as VGG-16, trained to perform invariant object recognition (Simonyan & Zisserman, 2014). (Note that this encoding is not foveated. Despite the importance of peripheral vision for understanding many relevant perceptual phenomena, for this mini-experiment we use an encoding that does not depend upon distance from the point of gaze.) We then took a set of arrays of 8 colored squares (Figure 8A), and measured the distance between them in the VGG-16 perceptual encoding space. This gives us a measure of the uncertainty in pinpointing images in perceptual encoding space. Humans have difficulty distinguishing these arrays in a VWM task, suggesting they have at least this much uncertainty. Given that same uncertainty, how well could we instead pinpoint a natural scene? Figure 8B shows scenes with approximately the same discriminability in VGG-16 space as the VWM displays. Figure 8C shows scenes that are more discriminable from this set.

The first thing to note from these results is that a distance metric applied to the last fully connected layer of VGG-16 seems to give us a reasonable measure of perceptual similarity. It is difficult to distinguish the arrays of colored squares from each other, and it is similarly difficult to distinguish the scenes in Figure 8B. The camera angle has changed somewhat, and the location of the number of vehicles and pedestrians has changed. The scenes in Figure 8C are more readily discriminated from those in Figure 8B. So the mini-experiment is a good first attempt. More importantly, note that for the same amount of uncertainty that makes an 8-item VWM task hard, one can pinpoint a scene fairly well. In a plausible perceptual encoding space, the same precision can specify either an array of about 8 items of random color and position, or mostly determine the scene, plus or minus some small changes. This suggests there is real hope for a unified explanation. The same inference limits that make VWM difficult allow a rich subjective experience of the real world. There is no need for information available in the rich subjective experience to be perversely unavailable for decision-making and action.

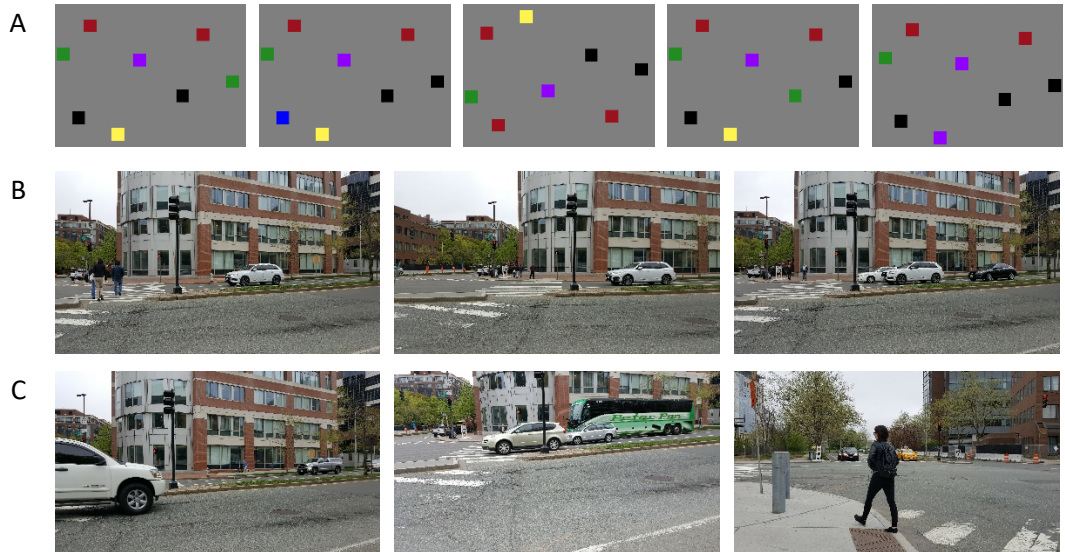

Figure 8. Rows A and B have similar mean discriminability in the perceptual encoding space of convolutional neural network VGG-16, trained on an object recognition task. Images in C are more discriminable from those in B.

## Conclusions

I have argued that perception results from limited inference processes, acting upon a perceptual encoding that has lost information, particularly from the peripheral visual field (Rosenholtz, 2017). A dominant source of lost information derives from encoding the visual input in terms of a rich set of summary image statistics. That perception results from inference suggests that there is some truth to the "illusion" theories of awareness. One perceives the results of inference, not some image projected in the head and viewed by the self. Perception is inherently something of an illusion. However, the illusion is not as extreme as previously thought, because vision is not so impoverished. Rather, tasks that seem to show impoverished vision are simply harder tasks than getting a rich impression of the world. Looked at in the right way, with the right model in hand, there is no awareness puzzle.

## Acknowledgements

The work described here was funded in part by NIH-NEI EY021473 and NIH NEI R21-EY019366 to Ruth Rosenholtz, and NSF/BMBF IIS-1607486 to Ruth Rosenholtz and Christoph Zetzsche. Thanks to Shaiyan Keshvari and Yrvine Thelusma for help with the VGG-16 experiment, and to Benjamin Wolfe and Shaiyan Keshvari for useful discussions.

## References

Alvarez, G. A. (2011). Representing multiple objects as an ensemble enhances visual cognition. *Trends. Cogn. Sci., 15,* 122-131.

Ariely, D. (2001). Seeing sets: Representation by statistical properties. *Psychol. Sci., 12*, 157-162.

Balas, B.J., Nakano, L. & Rosenholtz, R. (2009). A summary-statistic representation in peripheral vision explains visual crowding. *J. of Vision, 9*(12):13.

Blackmore, S. J., Brelstaff, G., Nelson, K., Troscianko, T. (1995). Is the richness of our visual world and illusion? Transsaccadic memory for complex scenes. *Perception, 24,* 1075-1081.

Block, N. (2011). Perceptual consciousness overflows cognitive access. *Trends in Cognitive Sciences, 15*(12), 567-575.

Chang, H. & Rosenholtz, R. (2016). Search performance is better predicted by tileability than by the presence of a unique basic feature. *J. of Vision, 16*(10):13.

Cohen, M. A., Dennett, D. C., & Kanwisher, N. (2016). What is the bandwidth of perceptual experience? *Trends in Cognitive Sciences, 20*(5), 324-335.

Dennett, D. C. (1991). *Consciousness explained.* Boston: Little Brown.

Dennett, D. C. (1998). No bridge over the stream of consciousness. *Behavioral and Brain Sciences, 21,* 753-754.

Ehinger, K. A. & Rosenholtz, R. (2016). A general account of peripheral encoding also predicts scene perception performance. *J. of Vision, 16*(2):13.

Freeman, J. & Simoncelli, E. P. (2011). Metamers of the ventral stream. *Nature Neuroscience, 14*(9), 1195-1201.

Haberman, J. & Whitney, D. (2009). Seeing the mean: Ensemble coding for sets of faces. *J. Exp. Psychol. Hum. Percept. Perf. 35*, 718-734.

Haberman, J. & Whitney, D. (2011). Ensemble perception: Summarizing the scene and broadening the limits of visual processing. *A Festschrift in honor of Anne Treisman (Wolfe, J. & Robertson, L., eds).*

Keshvari, S. & Rosenholtz, R. (2016). Pooling of continuous features provides a unifying account of crowding. *J. of Vision, 16*(3):39.

Lamme, V. (2010). How neuroscience will change our view on consciousness. *Cognitive Neuroscience, 1,* 204-220.

Leib, A., Kosovicheva, A., & Whitney, D. (2016). Fast ensemble representations for abstract visual impressions. *Nature Communications, 7*, 13186.

Luck, S. J. & Vogel, E. K. (1997). The limited capacity of visual working memory for features and conjunctions. *Nature, 390,* 279-281.

Mack, A. & Rock, I. (1998). Inattentional blindness. Cambridge, MA: MIT Press.

Oliva, A. & Torralba, A. (2006). Building the gist of a scene: The role of global image features in recognition. *Progress in Brain Research, 155,* 23-36.

O'Regan, J. K. (1992). Solving the "real" mysteries of visual perception: The world is an outside memory. *Canadian Journal of Psychology, 46,* 461-488.

Rensink, R. A., O'Regan, K., & Clark, J. (1997). To see or not to see: The need for attention to perceive changes in scenes. *Psychological Science, 8*(5), 368-373.

Rosenholtz, R. (2011). What your visual system sees where you are not looking. *Proc. SPIE 7865, Hum. Vis. Electron. Imaging, XVI, San Francisco, Feb. 2.*

Rosenholtz, R. (2016). Capabilities and limitations of peripheral vision. *Annual Rev. of Vision Sci., 2*(1), 437-457.

Rosenholtz, R. (2017). Capacity limits and how the visual system copes with them. *J. of Imaging Science and Technology (Proc. HVEI, 2017)*.

Rosenholtz, R. Huang, J., & Ehinger, K. A. (2012a). Rethinking the role of top-down attention in vision: Effects attributable to a lossy representation in peripheral vision. *Front. Psychol. 3*:13.

Rosenholtz, R., Huang, J., Raj, A., Balas, B. J., & Ilie, L. (2012b). A summary statistic representation in peripheral vision explains visual search. *J. of Vision, 12*(4):14.

Scimeca, J. M. & Franconeri, S. L. (2015). Selecting and tracking multiple objects. *Wiley Interdisciplinary Rev. Cogn. Sci. 6,* 109-118.

Simonyan, K. & Zisserman, A. (2014). Very deep convolutional networks for large-scale image recognition. *arXiv Technical Report, arXiv:1409.1556 (2014).*

Treisman, A. & Gelade, G. (1980). A feature-integration theory of attention. *Cogn. Psych. 12,* 97-136.

VanRullen, R., Reddy, L., & Koch, C. (2004). Visual search and dual tasks reveal two distinct attentional resources. *J. Cogn. Neurosci. 16,* 4-14.

Yamins, D. L. K., Hong, H., Cadieu, C. F., Solomon, E. A., Seibert, D., & DiCarlo, J. J. (2014). Performance-optimized hierarchical models predict neural responses in higher visual cortex. *PNAS, 111,* 8619-8624.

Zhang, X., Huang, J., Yigit-Elliott, S., & Rosenholtz, R. (2015). Cube search, revisited. *J. of Vision, 15*(3):9.